\documentclass[pra, a4paper, aps, twocolumn, showpacs,superscriptaddress, groupedaddress]{revtex4}
\usepackage[dvips]{graphicx}
\usepackage{ae}
\usepackage[T1]{fontenc}
\usepackage[ansinew]{inputenc}
\usepackage{amsmath}
\usepackage{amssymb}
\usepackage{graphicx}
\usepackage{color}
\usepackage[colorlinks]{hyperref}
\usepackage{lscape}
\begin{document}
\title{Revealing Advantage in a Quantum Network}
\author{Kaushiki Mukherjee}
\email{kaushiki_mukherjee@rediffmail.com}
\affiliation{Department of Mathematics, Dr. A. P. J. Abdul Kalam Government College, Kolkata, India.}
\author{Biswajit Paul}
\email{biswajitpaul4@gmail.com}
\affiliation{Department of Mathematics, South Malda College, Malda, West Bengal, India.}
\author{Debasis Sarkar}
\email{dsappmath@caluniv.ac.in}
\affiliation{Department of Applied Mathematics,  University of Calcutta,  92,  A.P.C. Road,  Kolkata-700009, India.}

\begin{abstract}
The assumption of source independence was used to reveal nonlocal (apart from standard Bell-CHSH scenario) nature of correlations generated in entanglement swapping experiments. In this work, we have discussed the various utilities of this assumption to reveal nonlocality (via generation of nonbilocal correlations) and thereby exploiting quantumness under lesser requirements compared to some standard means of doing the same. We have also provided with a set of sufficient criteria, imposed on the states(produced by the sources) under which source independence can reveal nonbilocal nature of correlations in a quantum network.
\end{abstract}
\date{\today}
\pacs{03.65.Ud, 03.67.Mn: Keywords: Nonlocality, Nonbilocality, Bell inequality}
\maketitle
\section{Introduction}
The analysis of correlated statistics of measurement outcomes in a quantum network has emerged as a recent trend in the study of understanding correlations\cite{BRA,BRAN,Fritz,Ross,Ros,Tav,km}. This sort of analysis has enriched the study of quantum theory from both theoretical as well as application viewpoint. From the theoretical viewpoint, this study led to the discovery of nonlocal nature of quantum predictions that cannot be described by any locally causal model \cite{Bell}. As applications, this has revealed the potential of nonlocal nature of certain quantum states, used in quantum information technologies, specifically for private randomness generation \cite{Pironio,Colbeck}, quantum key distribution \cite{Acin,Mayer}, for reducing communication complexity \cite{Cleve}, device-independent entanglement witnesses \cite{Bancal}, device-independent quantum state estimation \cite{Bar,Mck}. The key feature of all these applications is the fact that quantum non locality could be used in a device independent manner. Besides, this study can also be applied in various other tasks involving quantum networks such as quantum distributed computing \cite{BRAN}, quantum repeaters \cite{San,HAM,Zuk}. \\
Entanglement swapping experiments \cite{Zuk} reveal nonlocal nature of certain class of  quantum states (entangled states) more strongly compared to standard Bell- CHSH scenario\cite{Cl}. Motivated by this intuition, \textit{bilocal scenario} was introduced in \cite{BRA} and further analyzed in \cite{BRAN,Fritz,Ross,Ros,Tav,km}. \textit{Bilocal scenario} is a general scenario of quantum network involving three parties characterized by independence of two sources shared between the parties. This concept of source independence has emerged as a step towards exploiting nonlocal character of quantum correlations in a more common sense than that can be usually thought in a usual Bell scenario, thereby revealing quantum advantage in a network. For instance, a visibility of \textit{V} $> 50\%$ is enough to reveal quantumness (nonbilocality) in a \textit{bilocal} network using Werner state\cite{Wer} in contrast to a visibility of \textit{V}$> 70.7\%$(for projective measurements\cite{Wer})and \textit{V}$> 66\%$ approximately(for POVM \cite{Barr}), required for a Werner state to be nonlocal in usual Bell sense. This in turn guaranteed presence of local but nonbilocal correlations.  \\
In this context, it was interesting to observe that the assumption of bilocality can be applied to lower down the requirements to demonstrate quantumness in a system compared to some preexisting standard procedures. For instance,   sequential measurements are usually required to be performed on some states in order to reveal the hidden nonlocality of the states(\cite{pop}). Interestingly we have  considered  a particular family of X states(\cite{Hir}) which reveals hidden nonlocality under a sequence of suitable measurements. But the same state, when violates bilocal assumption in a network, generates nonbilocal and hence nonlocal(apart from standard CHSH sense)in absence of any sequence of measurements thus showing the efficiency of this assumption to reveal nonlocality compared to sequential measurements. In order to provide another instance we consider the concept of steering which is considered as failure of a hybrid LHV(local hidden variable)-LHS(local hidden state) model to generate the correlations between two subsystems(\cite{Wise, nat}). From set theoretic viewpoint, set of steerable states forms a subset of entangled states. So there exist some local entangled states which exhibit this weaker form of nonlocality. Again nonbilocality can also be considered a weaker form of nonlocality. This in turn gives rise to the intuition that there is a possibility of existence of non steerable states whose nonlocal nature can be guaranteed in a bilocal scenario.
\subsection{Summary of our work}
In this paper, we have derived the bound for violation of the bilocal inequality(Eq.\ref{A3}) by X states\cite{tyu} under projective measurements. This bound in turn has helped us to prove the efficiency of the source independence assumption to exploit nonlocality of an X state over some standard means. Specifically we have analyzed generation of local but nonbilocal correlations in a network using T states\cite{rho} which form a subclass of X states. The bilocal inequality may also be used to reduce the requirement of filtering operations(particular type of sequential measurements) over some X states to reveal hidden nonlocality. Also we    provided with example of some quantum states which may not be steerable both before and after being swapped in a network but can generate nonbilocal correlations. In totality, we have shortly discussed the various possibilities of revealing nonlocal(nonbilocal)correlations that the bilocal assumption provides in a quantum network. Now from that perspective one obvious question arises: \textit{what may be the criteria to be satisfied by a quantum state to produce nonbilocal correlations in a network? } The last part of our work basically focusses on this query where we have tried to make a systematic study of the constraints which when imposed on a two qubit state suffice to produce nonbilocal correlations in a network. From set theoretic  viewpoint  the set of bilocal correlations forms a subset of the set of  local correlations\cite{Ross}. Hence local correlations obtained by using two local copies may be nonbilocal.  Interestingly, by using the bilocal inequality(Eq.(\ref{A3}))to detect nonbilocal correlations, no two local copies of an X state, if used in a bilocal network can generate nonbilocal correlations. But the bilocal inequality(Eq.(\ref{A3}))being only a sufficient criteria for detection of nonbilocality, no such conclusion can be made in general. In \cite{BRA} Branciard et.al. gave an example showing that nonbilocal correlations can be generated in a quantum network using a separable state along with an entangled state. Intuitively certain amount of entanglement of atleast one of the two copies used is the minimal requirement for generation of nonbilocal correlations in a quantum network. However, in this manuscript we have analyzed the restrictions to be maintained by the copies of quantum states to generate nonbilocal correlations in a network via violation of the bilocal inequality(Eq.(\ref{A3})). Due to obvious complications arising in the analytic study for any two qubit state, we have chosen the class of T states and have given a set of criteria which when imposed on two copies of T states suffice to generate nonbilocal correlations in a quantum network.\\
 Our choice to consider X states only is supported by wide availability of X states in experimental works. The class of X states is of particular interest as it includes many well known class of states such as Bell diagonal states, Werner states(a one-parameter  family of states which encompasses both separable and entangled states\cite{Wer}), etc \cite{BEO}. In fact, any two qubit state can be converted to a X state by passing it through a noisy channel\cite{deo}. Besides, density matrix structure of two spin X states are used in many physical scenarios and can also be achieved in various experiments \cite{CHI,LDI,PET,PRA,BOS,WAN,HAG}. For instance, this class of states was encountered in \cite{Ceo} while analyzing entanglement of an atom in a quantized electromagnetic field. They constitute the spectra of all the systems with odd-even symmetry, for example, in the Ising and the XY models\cite{OSB,OST}. X states were also studied in condensed matter systems and in various other fields of quantum mechanics\cite{FAN,DIL,SAR,WER,CIL}.  The evolution of entanglement in X states(subjected to spontaneous emission) was analyzed in \cite{TYU} where it was further shown that some forms of X states remain invariant under general decoherence \cite{TYU} whereas some disentangle within finite time. In \cite{VIN}, the author showed that the  “sudden death of entanglement” of this class of states can be increased, decreased or averted with the aid of local operations. Thus X states emerge as an important two qubit class of states. Here we analyze the restrictions to be imposed on the state parameters of two copies of an X state to generate nonbilocal nature of correlations when used in a quantum network.\\
The paper may be broadly divided into the following sections: section II deals with some mathematical prerequisites, section III deals with the various directions in which bilocal assumption in a quantum network emerges as a better mean of demonstrating nonlocality(nonbilocality) compared to some standard methods of doing so. In section IV we discuss about requirement of entanglement for generation of nonbilocal correlations when two same copies of T states are used. In section V we give the set of sufficient criteria for generation of nonbilocal correlations, in forms of constraints on state parameters of two copies of T states.

\section{Preliminaries}
\subsection{Bilocal Scenario of Two-Qubit States}
Consider first the bilocal experimental setup (FIG.1)\cite{BRA,BRAN}. There are three parties Alice($A$), Bob($B$) and Charlie($C$) arranged in a linear pattern such that any pair of adjacent parties share a source and the two sources $S_1$ and $S_2$ are independent to each other. Each of these two sources $S_1$ and $S_2$ sends a physical system represented by $\lambda_1$ and $\lambda_2$ respectively. Independence of $S_1$ and $S_2$ guarantees independence of the two variables. All parties can perform measurements on their systems labeled by $x,\,y,\,z$ for Alice, Bob and Charlie and they obtain outcomes denoted by $a,\,b,\,c$ respectively. In particular, Bob might perform a joint measurement on the two systems that he receives from the two independent sources. The correlations obtained thereby take the form:

$p(a, b, c|x, y, z)=\iint d\lambda_{1} d\lambda_{2} {\rho(\lambda_1,\lambda_2)}$
\begin{equation}\label{p11}
P(a|x, \lambda_1)P(b|y, \lambda_1, \lambda_2)P(c|z, \lambda_2)
\end{equation}
where $\lambda_1$ characterizes the joint state of the system produced by the source $S_1$ and $\lambda_2$ for the system $S_2$. The hidden states $\lambda_1, \lambda_2$ follow independent probability distribution $\rho_1(\lambda_1)$ and $\rho_{2}(\lambda_2)$ such that
\begin{equation}\label{p2}
    \rho(\lambda_1,\lambda_2)=\rho_1(\lambda_1)\rho_2(\lambda_2)
\end{equation}
\begin{figure}[htb]
\centering
\includegraphics[width=2.5in]{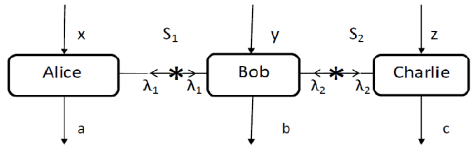}
\caption{\emph{The bilocal scenario(\cite{BRA,BRAN}) where three parties Alice,  Bob and Charlie  share two sources $S_1$ and $S_2$(characterized by the hidden states $\lambda_1$ and $\lambda_2$ respectively) which are assumed to be independent.}}
\end{figure}
Clearly one can consider entanglement swapping procedure (with independent sources)\cite{BRA,BRAN,Fritz,Ross,Ros} as a particular case of this scenario. The correlations of the form (Eq.(\ref{p11})) and (Eq.(\ref{p2})) satisfy the inequality $\textbf{B}\leq1$(\cite{BRAN}) where:
\begin{equation}\label{A3}
   \textbf{B}= \sqrt{|I|} + \sqrt{|J|}
\end{equation}

$\textmd{with}\,\,I$=$\frac{1}{4}\sum \limits_{x, z=0,1}\langle A_x B^0 C_z\rangle,\,J$=$\frac{1}{4}\sum \limits_{x, z=0, 1}(-1)^{x+z}\langle A_x B^1 C_z\rangle\,$ $\textmd{and}\,\langle A_x B^y C_z\rangle$=$\sum\limits_{a, b^0b^1, c}(-1)^{a+b^y+c}P(a, b^0b^1, c|x, z).$ Here $A_x$ and $C_z$ are the observables for binary inputs $x,\,z$ of Alice and Charlie respectively whereas $B^ y$ denotes the observable of Bob corresponding to a single input. For Alice and Charlie there are two outputs $a,c\in\{0,1\}$ respectively whereas for Bob there are four outputs labeled by two bits $\overrightarrow{\textbf{b}}=b^0b^1=00,\,01,\,10,\,11.$
\subsection{X states}
The class of X states\cite{tyu} is given by:
$$\chi\,=\,\varsigma|00\rangle \langle 00|+\kappa|01\rangle\langle 01|+\zeta|10\rangle\langle10|$$
\begin{equation}\label{st9}
+d|11\rangle\langle11|+p(|00\rangle\langle11|+|11\rangle\langle00|)+q(|01\rangle\langle10|+|10\rangle\langle01|)\quad \end{equation}
with $ p$ and $q$ real, $\varsigma+\kappa+\zeta+d=1$. $\chi\geq0$ demands $p^ 2\leq \varsigma d$ and $q^ 2\leq \kappa\zeta$. The corresponding state parameters of two different copies of an X state, $\chi_i(i=1,2)$ are thus given by $\varsigma_i$, $\kappa_i$, $\zeta_i$, $d_i$, $p_i$ and $q_i(i=1,2)$ respectively. Any X state can also be represented in density matrix formalism as\cite{ali}:
\begin{equation}\label{x}
\chi_i=
   \left(\begin{array}{cccc}
      \varsigma_i &0 & 0 & p_i\\
      0 & \kappa_i & q_i & 0\\
      0 & q_i &\zeta_i & 0\\
      p_i &0 &0 & d_i \\
       \end{array} \right),
\end{equation}
$\,i=1,2\, \textmd{for two copies}$.\\
 \textit{T states} \cite{rho}: This class of states forms a subclass of X states and has maximally mixed marginals:
 \begin{equation}\label{p1}
    \tau_i= \frac{1}{4}\left(\begin{array}{cccc}
      1+c_{3i} &0 & 0 & c_{1i}-c_{2i}\\
      0 & 1-c_{3i} & c_{1i}+c_{2i} & 0\\
      0 & c_{1i}+c_{2i} &  1-c_{3i} & 0\\
      c_{1i}-c_{2i} &0 &0 & 1+c_{3i} \\
       \end{array} \right)
 \end{equation}
 where $|c_{ji}|\leq1\,(j=1,2,3)$ and $|c_{1i}\pm c_{2i}|<1\mp c_{3i},\,(i=1,2)$. Clearly for this subclass, $\varsigma_i=d_i, \,\kappa_i=\zeta_i\,(i=1,2).$ This class of states is equivalent to Bell mixture. Hence the well known class of Werner states belong to this class of X states.

\subsection{Steering nonlocality}
The concept of steering may be considered as a signature of non-classicality of quantum correlations\cite{zuk1}, intermediate between entanglement and nonlocality. Bell inequalities are used to detect correlations compatible with local
causal models. For two parties Alice and Bob(say), the corresponding correlations satisfy:
\begin{equation}\label{st1}
    p(a, b|\overrightarrow{x}, \overrightarrow{y})=\int d\lambda{\rho(\lambda)}P(a|\overrightarrow{x}, \lambda)P(b|\overrightarrow{y}, \lambda)
\end{equation}
where $a$, $b$ denote local results, $\overrightarrow{x},\overrightarrow{y}$ the corresponding local inputs and $\lambda$ denotes the hidden variable. The bipartite correlations are non steerable if they satisfy the following model \cite{zuk1}:
\begin{equation}\label{st2}
  p(a, b|\overrightarrow{x}, \overrightarrow{y})=\int d\lambda{\rho(\lambda)}P(a|x, \lambda)\textmd{Tr}[\widehat{\Pi}(b|\overrightarrow{y})\rho_2(\lambda)]
\end{equation}
where $\widehat{\Pi}(b|\overrightarrow{y})$ is the projection operator corresponding to an observable
characterized  by Bob's setting $\overrightarrow{y}$, associated with the eigenvalue $b$ and $\rho_2(\lambda)$ corresponds to some pure state of Bob’s system , parameterized by variable $\lambda$. There exist states which do not violate any Bell inequality but are steerable in nature(\cite{Wise}). In \cite{zuk1} Zukowski et.al. gave a criteria sufficient to detect steerability of a bipartite quantum state. The key feature of their criteria is that if  $\overrightarrow{u}$ and $\overrightarrow{v}$(members of an arbitrary Hilbert space) are such that $||\overrightarrow{u}||^2>|\langle\overrightarrow{u}|\overrightarrow{v}\rangle|$ then surely $\overrightarrow{u}\neq \overrightarrow{v}.$ For their purpose they considered quantum correlation function $E_{QM}$ and non steerable correlation function(local hidden state correlations)$E_{NS}$. $E_{QM}$ is defined as:
\begin{equation}\label{st3}
   E_{QM}\equiv E_Q(\overrightarrow{m},\overrightarrow{n} )=   \textmd{Tr}[\overrightarrow{m}.\overrightarrow{\sigma}\bigotimes\overrightarrow{n}.\overrightarrow{\sigma}\rho_{12}]
\end{equation}
$\rho_{12}$ is the bipartite quantum state shared between the two parties. In general the state density matrix can be decomposed as:
\begin{equation}\label{st4}
    \rho_{12}=\frac{1}{2^2}\sum_{i_1,i_2=0}^{3}t_{i_1i_2}\sigma^1_{i_1}\bigotimes\sigma^2_{i_2}
\end{equation}
 where $\sigma^k_0,$ denotes the identity operator in the Hilbert space of qubit k and $\sigma^k_{i_k},$ are the Pauli operators along three perpendicular directions, $i_k=1,2,3$.
 The components $t_{i_1i_2}$ are real and given by $t_{i_1i_2}=\textmd{Tr}[\rho_{12}\sigma^1_{i_1}\bigotimes\sigma^2_{i_2}]$. Using Eq.(\ref{st4}), $E_{QM}$ becomes
 \begin{equation}\label{st5}
    E_Q(\overrightarrow{m},\overrightarrow{n} ) =\sum_{i_1,i_2=1}^{3}t_{i_1i_2}m_{i_1}n_{i_2}
 \end{equation}
 $m_{i_1}$ and $n_{i_2}$ denote the Cartesian coordinates of the bloch vectors $\overrightarrow{m}$ and $\overrightarrow{n}$ defining corresponding measurement alignments and $t_{i_1i_2}(i_1,i_2=1,2,3)$ denoting the elements of the correlation tensor $t$ corresponding to the density matrix $\rho_{12}.$ The non steerable correlation function($E_{NS\backslash LHS}$) can be defined as(\cite{zuk1}):
 \begin{equation}\label{st6}
    E_{NS}\equiv E_{NS}(\overrightarrow{m},\overrightarrow{n})= \sum_{\lambda}\rho(\lambda)I(\overrightarrow{m},\lambda)Q(\overrightarrow{n},\lambda)
 \end{equation}
 with $I(\overrightarrow{m},\lambda)=P(1|\overrightarrow{m},\lambda)-P(-1|\overrightarrow{m},\lambda)$ and $Q(\overrightarrow{n},\lambda)=\textmd{Tr}[\overrightarrow{n}.\overrightarrow{\sigma}\rho_2({\lambda})]$, $\rho_2({\lambda})$ being the quantum particle on Bob's side. In \cite{zuk1}, Zukowski et.al. showed that the shared quantum state($\rho_{12}$) is steerable(from Alice to Bob) if the corresponding correlation functions satisfy the Bell-type inequality:
   \begin{equation}\label{st7}
    \langle E_{QM}|E_{QM}\rangle>\langle E_{QM}|E_{LHS}\rangle.
   \end{equation}

   They used this basic geometric approach to design the sufficient criteria for steering:
   \begin{equation}\label{st8}
    ||t||_{\infty}<\frac{2}{3}||t||^2
   \end{equation}
   where $||t_{\infty}||$ and $||t||$ denote spectral norm and Hilbert-Schmidt norm  of $t$ respectively. Any quantum state which satisfies this criteria(\ref{st8}) is steerable in nature. They deduced(using Eq.(\ref{st5})) the more simpler form of the criteria as:
\begin{equation}\label{st9i}
\textmd{Max}_{\overrightarrow{m},\overrightarrow{n}}\{E(\overrightarrow{m},\overrightarrow{n})\}\leq\sum_{i,j=1}^3t_{ij}^2.
\end{equation}
If any state violates the steering criteria given by Eq.(\ref{st9i}) then the steerability of the state cannot be guaranteed by this geometric approach.
\subsection{Hidden nonlocality}
 Apart from standard Bell tests, nonlocality in a bipartite quantum state can  be revealed by subjecting the system to a sequence of measurements. This was first shown by Popescu\cite{pop} who argued that $d$ dimensional Werner class of states($d\geq 5$) can show nonlocality if each of the two  parties can perform a sequence of two projective measurements though some entangled states of this class cannot reveal nonlocality in standard Bell-CHSH sense. This type of nonlocality was referred to as hidden nonlocality\cite{pop,mas1,mas2}. A particular type of sequential measurements is given by filtering operations. If $\rho_{AB}$ denotes a state shared between two parties, Alice and Bob(say) and $F_{A},\, F_{B}$ denote filtering operations(projective measurements) performed by Alice and Bob on their particles respectively, then the final(filtered) state is given by
 \begin{equation}\label{r1}
  \hat{\rho}_{AB}=\frac{1}{N}(F_A\bigotimes F_B \rho_{AB}F_A^{\dag}\bigotimes F_B^{\dag})
\end{equation}
where $N$ is the normalization factor. If $\hat{\rho}_{AB}$ violates a Bell inequality then $\rho_{AB}$ is said to reveal hidden nonlocality.
\subsection{Horodecki criteria}
 Let$\rho_{12}$ be any bipartite state. Let
\begin{equation}\label{s}
M(\rho_{12})=\sqrt{t_{11}^2+t_{22}^2}
\end{equation}
where $t_{11}^2$ and $t_{22}^2$ are the two largest eigen values of $t^Tt$ where $t^T$ is the transpose of the correlation tensor $t$ of $\rho_{12}$(Eq.(\ref{st4})).
By Horodecki criteria \cite{HOR}, a state is nonlocal in the sense of Bell-CHSH\cite{Cl}, if
\begin{equation}\label{h}
M(\rho_{12})>1
\end{equation}
For X states,
\begin{equation}\label{x3i}
\chi_i=\textmd{Max}_{k}\{\Theta_{i,k}\},\,k=1,2,3\, \textmd{and} \,i=1,2
\end{equation}
where
\begin{itemize}
\item $\Theta_{i,1}= 8(p_i^2+q_i^2)$
\item $\Theta_{i,2}= (\varsigma_i-\kappa_i-\zeta_i+d_i)^2+4(p_i+q_i)^2$
\item $\Theta_{i,3}= (\varsigma_i-\kappa_i-\zeta_i+d_i)^2+4(p_i-q_i)^2$
\end{itemize}
We now introduce variables $\epsilon_i$, $\delta_i$ and $\xi_i(i=1,2)$ such that:
\begin{eqnarray}\label{x4ia}
\Theta_{i,1} &=& 1-\epsilon_i \\
\Theta_{i,2} &=& 1-\delta_i\\
\Theta_{i,3} &=& 1- \xi_i
\end{eqnarray}
where i=1,2 for two copies of X state $\chi_1$ and $\chi_2$. If $\chi_i$ is local then each of $\epsilon_i$, $\delta_i$ and $\xi_i$ must lie in [0,1] whereas if nonlocal then at least one of these three variables must lie in [-1,0).
\section{Advantage Of Bilocal Assumption In a Network}
In a bilocal network, let source $S_1$ sends one copy of X state $\chi_1$(\ref{st9}) to Alice and Bob and  source $S_2$ sends a different copy of X state $\chi_2$ to Bob and Charlie. Let Bob first performs full Bell basis measurement on his two particles, where the four outcomes being labeled as $b^0b^1=00,01,10,11$. Outputs are designated as $00,01,10,11$, if he obtains Bell states $|\phi^+\rangle$, $|\phi^-\rangle$, $|\psi^+\rangle$ and $|\psi^-\rangle$ respectively. The final state is shared between Alice and Charlie. Each of these two parties then perform any of two projective measurements. Under arbitrary projective measurement settings for Alice and Charlie, the maximum value of the bilocal operator$\textbf{B}$(Eq.\ref{A3}), obtained by maximization over measurement angles, is given by(see Appendix A):
\begin{equation}\label{rii}
    \textbf{B}\leq B_1,
\end{equation}
where $$B_1= \sqrt{\Pi_{i=1}^2(\varsigma_i-\kappa_i-\zeta_i+d_i)+4|\Pi_{i=1}^2(p_i+q_i)|}.$$ Hence the bilocal inequality (Eq.(\ref{A3})) takes the form:
\begin{equation}\label{r5}
    \sqrt{\Pi_{i=1}^2(\varsigma_i-\kappa_i-\zeta_i+d_i)+4|\Pi_{i=1}^2(p_i+q_i)|}\,\leq\,1
\end{equation}
Violation of this inequality(Eq.\ref{r5}) by any two copies of X state guarantees that the correlations generated by using these two copies in a quantum network( characterized by source independence) are nonbilocal in nature.
\subsection{Local but nonbilocal correlations}
In a network of three parties, a necessary condition for any tripartite correlation to be local \cite{Pir} is that the corresponding  bipartite correlations shared between Alice and Charlie(conditioned on a particular output of Bob) must satisfy CHSH inequality \cite{Cl}. If two copies of T states(\ref{p1}) are used in a bilocal network, as depicted above, the bipartite correlations in between Alice and Charlie are local if these correlations satisfy,
\begin{equation}\label{r6}
\textmd{ Max}_{i,j\in\{1,2,3\},i\neq j}[\sqrt{c_{i1}^2c_{i2}^2+c_{j1}^2c_{j2}^2}]\leq 1
\end{equation}
 However violation of Eq.(\ref{r5}) guarantees the nonbilocality of the correlations, i.e., if the correlations satisfy the condition:
\begin{equation}\label{r7}
    \sqrt{\Pi_{j=1}^2c_{1j}+\Pi_{j=1}^2c_{3j}}>1.
\end{equation}
 Clearly for any possible value of state parameters $c_{ij}$, left hand side of Eq.(\ref{r7}) has a larger value than that of Eq.(\ref{r6}). There exists a possible range of state parameters $c_{ij}$(see FIG.2) for which both Eq.(\ref{r6}) and Eq.(\ref{r7}) are satisfied which in turn  points out that the bilocal assumption renders advantage in  a quantum network to reveal nonlocality(nonbilocality).
  \begin{center}
\begin{figure}[htb]
\centering
\includegraphics[width=2.5in]{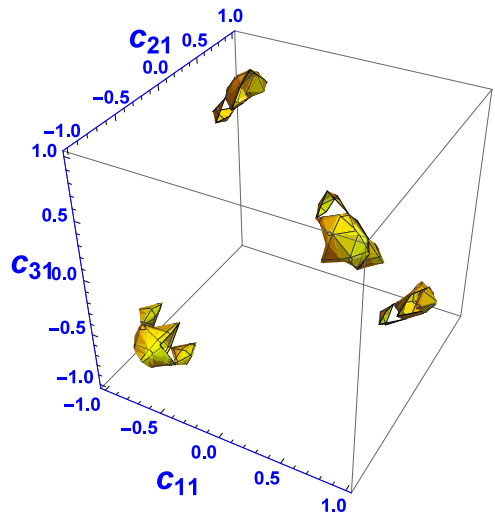}
\caption{\emph{The shaded region gives the region of local but nonbilocal correlations when two same copies of T states($\tau_1$) are used in a bilocal network.}}
\end{figure}
\end{center}
\subsubsection{Resistance to noise} The last subsection guarantees the existence of local but nonbilocal correlations in a network. It was already pointed out by Branciard.et.al.(\cite{BRAN}), that the most important physical interpretation of such correlations can be obtained by quantifying the resistance to noise of the nonbilocal correlations. For that we consider  an entanglement swapping scenario where each of two independent sources $S_i$ produces a noisy two qubit state:
\begin{equation}\label{x4iii}
\Lambda_i=\alpha_i |\psi^-\rangle\langle\psi^-|+(1-\alpha_i)\frac{\mathbf{1}}{4},\,(\textmd{where}\,\alpha_i\in[0,1](i=1,2))
\end{equation}
This family of states belong to the class of T states.
Here $\alpha_i$ denotes the visibility of $\Lambda_i(i=1,2)$ which is a measure of the resistance to noise given by the state. The smallest visibility( $\alpha_i$) for which $\Lambda_i$ is local(in standard CHSH scenario) is called the \textit{local visibility threshold}($V_i$)(\cite{BRAN}). Analogously, the smallest visibility for which the correlation produced in the network is bilocal  is called the \textit{bilocal visibility threshold}($V_{biloc}$). Now for the tripartite correlations produced in the network to be local, i.e., the bipartite correlations between Alice and Charlie conditioned on any possible output of Bob to violate Eq.(\ref{r6}), the corresponding smallest visibility is given by $V_{loc}$. Clearly quantum advantage is obtained if $V_{biloc}<V_{loc}$. $V_{biloc}<V_{loc}$ corresponds to generation of local but nonbilocal correlations in the network which has already been discussed above.
Let two copies of the noisy state(Eq.(\ref{x4iii})) are used in a bilocal network. The visibility of bilocal and local correlations produced in a network using the noisy states(Eq.(\ref{x4iii}))are determined by $\prod_{i=1}^2\alpha_i$\cite{BRAN}.  By using the bilocal criteria(Eq.(\ref{r7})) for the T states(\ref{p1}), it can be said that the correlations produced are nonbilocal if $\prod_{i=1}^2\alpha_i>\frac{1}{2}$.  Hence $V_{biloc}=\frac{1}{2}$\cite{BRAN}.  Now the tripartite correlations generated in the network correlations are local(\ref{r6}) if $\prod_{i=1}^2\alpha_i>\frac{1}{\sqrt{2}}$. Hence $V_{loc}=\frac{1}{\sqrt{2}}.$
Clearly $V_{biloc}<V_{loc}$. Now it will be interesting to investigate the nature of the states(Eq.(\ref{x4iii})) produced by the sources that suffice to give quantum advantage, specifically to violate the bilocal inequality(Eq.(\ref{r7})):
 \begin{itemize}
 \item At least one of the two copies must be nonlocal.
\item Without loss of generality, let $\Lambda_1$ be local and $\Lambda_2$ nonlocal. Let $\phi_i(i=1,2)$ denote the parameters quantifying the amount of deviation of the visibility parameter $\alpha_i$ of $\Lambda_i(i=1,2)$ from the local visibility threshold $V_i$. Hence $\alpha_1=\frac{1}{\sqrt{2}}-\phi_1,\quad \alpha_2=\frac{1}{\sqrt{2}}+\phi_2$. Nonbilocal correlations are produced for: $\phi_1-\phi_2+\sqrt{2}\phi_1\phi_2<0$. $\phi_1$ and $\phi_2$ can thus be interpreted as measure of locality of $\Lambda_1$ and nonlocality of $\Lambda_2$ respectively that suffice to generate nonbilocal correlations in a network using $\Lambda_1$ and $\Lambda_2$. It is clear from FIG.3 that $\phi_1\in(0,0.21)$ and hence $\alpha_1\in(0.497,0.707)$ which in turn implies that the local copy must be entangled in order to violate Eq.(\ref{A3}).
\end{itemize}
However, if both $\Lambda_1$ and $\Lambda_2$ are nonlocal then Eq.(\ref{r7}) is violated if:
$\phi_1+\phi_2+\sqrt{2}\phi_1\phi_2>0\quad \textmd{where}\, \alpha_1=\frac{1}{\sqrt{2}}+\phi_1,\, \alpha_2=\frac{1}{\sqrt{2}}+\phi_2.$ But this case is not noteworthy as the two copies being nonlocal, generation of nonbilocal correlations becomes quite obvious.

\subsection{Nonbilocality versus hidden nonlocality}
As already discussed before, there exist some states whose nonlocality is revealed only when subjected to sequential measurements.
For analyzing the efficiency of bilocality assumption against the requirement of sequential measurements for revealing nonlocality of a  quantum state, we consider filtering operations. Let the state($\rho_{AB}$) shared between two parties  be an X state(Eq.(\ref{st9})). The class of projective measurements(filtering operations) are given as:
\begin{equation}\label{r2}
    F_{A}=\lambda_1|0\rangle\langle0|+|1\rangle\langle1|,\, \lambda_1\in[0,1]
\end{equation}
\begin{equation}\label{r3}
    F_{B}=|0\rangle\langle0|+\lambda_2|1\rangle\langle1|,\, \lambda_2\in[0,1]
\end{equation}
The density matrix of the filtered state(Eq.(\ref{r2})) is given by:
\begin{equation}\label{r4}
\hat{\rho}_{AB}=\frac{1}{N_1}\left(
  \begin{array}{cccc}
    \varsigma\lambda_1^2\lambda_2^2 \ & 0 & 0 &  p \lambda_1\lambda_2 \\
    0 & \kappa\lambda_1^2 & q\lambda_1\lambda_2& 0 \\
    0 & \lambda_1^2\lambda_2^2 &  \zeta\lambda_2^2 & 0 \\
      p \lambda_1\lambda_2 & 0 & 0 &  d \\
  \end{array}
\right)
\end{equation}
where $N_1=\varsigma\lambda_1^2\lambda_2^2+\kappa\lambda_1^2+\zeta\lambda_2^2+d$.\\
Now in the bilocal network, let source $S_1$(say) generates a state which is local in standard sense(upto projective measurements) but is nonlocal under filtering operation and let $S_2$ generates a nonlocal state and let nonbilocal correlations be generated. In this case it can be said that the state which was local in standard sense and may not generate any better than any classical correlation unless subjected to suitable filtering operations, when used in a bilocal network(hence no sequential measurement), now help to generate nonbilocal correlations. This in turn points out the efficiency of bilocal assumption over hidden nonlocality. To be more precise bilocal assumption reduces the requirement of filtering operations over a state so that it may be used for generating non local correlations. The class of X states are local if Eq.(\ref{h}) holds. Under filtering operations(Eqs.(\ref{r2}),(\ref{r3})), the bound for violation of CHSH inequality by the filtered state(Eq.(\ref{r4})) is given in Table(\ref{table1})(see Appendix C). After extensive numerical tests, we conjecture that for any local copy of X state($\chi_1$) which violates CHSH after filtering, when used in a bilocal network along with a nonlocal copy of X state($\chi_2$), nonbilocal correlations are generated under some restrictions imposed on both $\chi_1$ and $\chi_2$. This in turn justifies our claim that bilocal inequality(Eq.(\ref{r5})) may sometimes reduce the requirements of filtering operations for generation of non local correlations. In support of our conjecture we consider a particular subclass of
 X state\cite{Hir}:
\begin{equation}\label{x5}
\varrho=\alpha|\psi^-\rangle\langle\psi^-|+\frac{(1-\alpha)}{2}(|00\rangle\langle00|+|01\rangle\langle01|),\alpha\in[0,1]
\end{equation}
A local model under projective measurements of this class of states (which is entangled for $\alpha>0$) exists for $\alpha\leq\frac{1}{2}$. However this state is local for $\alpha \leq\frac{1}{\sqrt{2}}$(Horodecki criteria). Under filtering operations(Eqs.(\ref{r2}),(\ref{r3})) with $\delta=\frac{\epsilon}{\sqrt{\alpha}}$, this class exhibits hidden nonlocality for any $\alpha\,>\,0$ as the bound for violation of the filtered state(Table.\ref{table1}, see Appendix C) is given by $2\sqrt{1+\alpha}$. Now let two copies of this subclass $\varrho_i(i=1,2)$ are used  in a network(characterized by source independence). Let $\alpha_1\leq\frac{1}{2},$ i.e., $\varrho_1$ has local model. Let $\varrho_1$ exhibits hidden nonlocality, i.e., $\alpha_1\in(0,\frac{1}{2}]$. Let $\varrho_2$ be nonlocal. Hence $\alpha_2>\frac{1}{\sqrt{2}}$. The correlations generated in the network are nonbilocal for $V_{biloc}(=\prod_{i=1}^2 \alpha_i)\,>\,\frac{1}{2}$. However the correlations produced in the network are local as $V_{loc}(=\prod_{i=1}^2 \alpha_i)\,\leq 1$.  Hence  nonbilocality can be generated without performing sequential measurements for a restricted range of visibilities of the two copies which is the same as that for the class of states(Eq.(\ref{x4iii})) (see FIG.3). Clearly at least one of the two copies must be nonlocal and if $\varrho_2$ be maximum nonlocal then the visibility ($\alpha_1$) of the local copy ($\varrho_1$) must be at least $0.5$.
\begin{figure}[htb]
\centering
\includegraphics[width=2.2in]{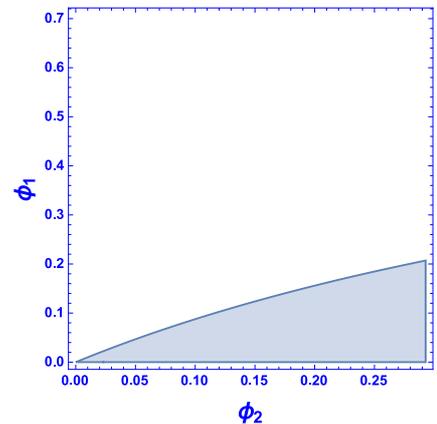}
\caption{\emph{This figure analyzes the tradeoff of the visibilities} $\alpha_1$\emph{and} $\alpha_2$ \emph{of the local} ($\varrho_1(\Lambda_1)$) \emph{and nonlocal} ($\varrho_2(\Lambda_2)$) \emph{copies} \emph{respectively in terms of their deviations from $V_i=\frac{1}{\sqrt{2}}.$} \emph{The restricted range of visibilities of the two copies that suffice to give quantum advantage when used in a network is given by } $\phi_1-\phi_2+\sqrt{2}\phi_1\phi_2<0$ \emph{where} $ \alpha_1=\frac{1}{\sqrt{2}}-\phi_1,\quad \alpha_2=\frac{1}{\sqrt{2}}+\phi_2.$ \emph{Clearly if one copy}  \emph{is maximum local then the other copy} \emph{suffices to be just nonlocal.}}
\end{figure}

\subsection{Quantumness of non steerable states}
The steering condition(Eq.(\ref{st9i})) and violation of inequality(Eq.(\ref{A3})) both can be considered as criteria that suffice to exploit non local behavior of quantum correlations. While the former signifies presence of weaker form of nonlocality in standard Bell-CHSH sense, violation of the latter is the signature of the presence of nonbilocality and hence nonlocality(apart from the standard Bell-CHSH notion). Here we mainly focus on comparing efficiency of these two inequalities to reveal nonlocality and consider $X$ states, for this task. Using the steering criteria(Eq.(\ref{st9i})) it becomes certain that any state(say $\rho$) of this class is steerable(from Alice to Bob's direction)if it satisfies:
\begin{equation}\label{st10}
\textmd{Max}\{|R_1|,|R_2|,|R_3|\}<\frac{2}{3}(R_1^2+R_2^2+R_3^2)
\end{equation}
where $R_1=2(p+q)$,$R_2=2(p-q)$ and $R_3=\varsigma-\kappa-\zeta+d$
Now let two identical copies of the state($\rho$) are used in an entanglement swapping network. Depending on any particular output of Bob's full basis measurement(say $|\psi^{+}\rangle$), the final state, shared in between Alice and Charlie, is steerable(from Alice to Charlie's direction) if it satisfies:
\begin{equation}\label{st11}
 \textmd{Max}\{|R^{'}_1|,|R_2^{'}|,|R_3^{'}|\}<\frac{2}{3}((R_1^{'})^2+(R_2^{'})^2+(R_3^{'})^2)
\end{equation}
with $R_1^{'}=\frac{2(p+q)^2}{W}$,$R_2^{'}=\frac{2(p-q)^2}{W}$ and $R_3^{'}=\frac{(\kappa+\zeta)R_3+2(\kappa\zeta-\varsigma d)}{W}$ and $W=(\kappa+\zeta)+2(\varsigma d-\kappa\zeta).$
If these two copies of $\rho$ are used in a bilocal network, where the intermediate party Bob performs full Bell basis measurement with the extreme two parties Alice and Charlie performing projective measurements, the correlations generated are nonbilocal in nature if :
\begin{equation}\label{st12}
    \sqrt{4R_1^2+R_3^2}> 1.
\end{equation}
There exist some states(see FIG.4) of this class which violate the steering criteria(Eq.(\ref{st9i})) both before and after being swapped in an entanglement swapping network. This is indicated when such a state($\rho$) initially violates Eq.(\ref{st10}) and the state, resulting  in the entanglement swapping network(using two identical copies of the state $\rho$) conditioned on any particular output of Bob, say $|\psi^{+}\rangle$ violates Eq.(\ref{st11}). Hence steerability of the state($\rho$) cannot be guaranteed when shared between two parties, both before and after being used in a network.  But when two copies of the state($\rho$) are used in a bilocal network, they generate nonbilocal correlations(Eq.(\ref{st12})). This in turn justifies our claim that bilocal criteria(Eq.(\ref{A3})) is more efficient in revealing non-classicality of a quantum system.

\begin{figure}[htb]
\centering
\includegraphics[width=2.2in]{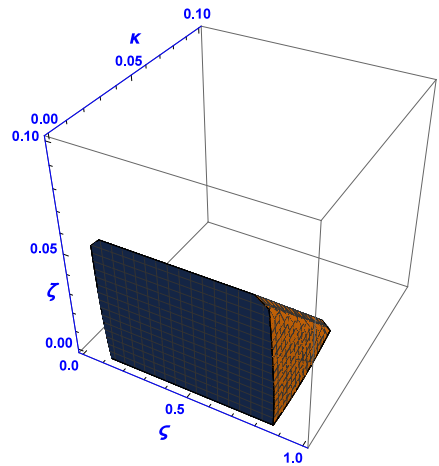}
\caption{\emph{This figure indicates a restricted region of the state parameters $\varsigma$, $\kappa$ and $\zeta$ of the class of X states(Eq.(\ref{st9}))for some fixed value of the other two state parameters($p=0.24$, $q=0$). A state, corresponding to any value of $\varsigma$, $\kappa$ and $\zeta$, lying in anyone portion of the shaded region, cannot be guaranteed to be steerable in nature both before and after being used in an entanglement network(considering the final state shared between Alice and Charlie is obtained if Bob obtains $|\psi^{+}\rangle$ as an output of full Bell basis measurement) but generates nonbilocal correlations if used in a bilocal network. This region in turn indicates the advantage of bilocal assumption over the concept of steering to exploit quantumness of a state.}}
\end{figure}
\section{Entanglement and nonbilocality}
Existence of entanglement in a bipartite state is necessary for the correlations produced by suitable measurements on the state in a quantum network to reveal nonlocality, not only in standard Bell-CHSH sense but also hidden nonlocality. In a bilocal network for generation of nonbilocal correlations via violation of the bilocal inequality(Eq.(\ref{A3})), two copies of the state used must be entangled. However the bilocal inequality(Eq.(\ref{A3})) is only a sufficient criteria for detection of nonbilocal correlations. As has been already discussed, Branciard et.al. gave an example in \cite{BRA} showing that the set of correlations generated in a network using a separable state along with an entangled state cannot have a bilocal decomposition(Eqs.(\ref{p11}),(\ref{p2})). However if one uses the bilocal inequality(Eq.(\ref{A3})) for detection of nonbilocality then that requires entanglement of both the copies of two qubit quantum states used in the network.  Here we have proved this analytically for two same copies of T states. Among several well-known measures of two qubit entanglement\cite{bnn,woo,vid} we pick the measure of concurrence\cite{woo}. Concurrence for the $T$ states reads as:
\begin{equation}\label{b50}
E(\tau_1)= \max\{0,\frac{|c_{11}-c_{21}|-|1-c_{31}|}{2},\frac{|c_{11}+c_{21}|-|1+c_{31}|}{2} \}
\end{equation}
$T$ states is separable if $E(\tau_1)=0$ i.e, $|c_{11}|+|c_{21}|+|c_{31}|\leq 1$. When two same copies of states from this family of states are used in the bilocal network,  nonbilocal correlations are generated if $c_{11}^2+c_{31}^2>1$. Now if it is possible to generate nonbilocal correlation by using two separable copies of $T$ states then $\sqrt{1-c_{11}^2}< 1-|c_{11}|-|c_{21}|$ i.e, when $|c_{21}|\leq 1- |c_{11}|-\sqrt{1-c_{11}^2}$. Now as $1- |c_{11}|\leq \sqrt{1-c_{11}^2}$, so the last inequality implies $|c_{22}|<0$. Hence entanglement is necessary to generate nonbilocal correlation by using two same copies of $T$ states. From extensive numerical tests, we can safely conclude that even if we use two different copies of T states, then both of them must be entangled so as to generate nonbilocal correlations. In fact we have argued in the next section that not even two different copies of local states can generate nonbilocal correlations under projective measurements.
\section{Sufficient Criteria Of Nonbilocality}
After analyzing the manifold utilities of the bilocal assumption for testing quantumness in a network, it is however interesting to note that there exist some nonbilocal correlations that satisfy the bilocal inequality(Eq.\ref{A3}) but violation of this inequality by any correlation ensures nonbilocality. Thus violation of this inequality can be regarded as a convenient tool for testing nonbilocality. So from the application viewpoint it is interesting to analyze the criteria under which the physical systems (state) sent by the sources exhibit nonbilocality. If source $S_i$ emits $\chi_i$$(i=1, 2)$ then bilocality equation takes the form (for proof, see Appendix B):
\begin{equation}\label{x4}
  \sqrt{\sqrt{\prod_{i=1}^2(1-\delta_i+\epsilon_i-\xi_i)}+\sqrt{\prod_{i=1}^2(1-\delta_i-\epsilon_i+\xi_i)}}\,\leq \sqrt{2}
\end{equation}
Under projective measurements, if both the copies of X state used in the network are local then Eq.(\ref{x4}) cannot be violated(see Appendix B).
Let us now assume one of the two copies used in the network be nonlocal. Let $\chi_1$, $\chi_2$ be the local and nonlocal copy respectively. Let $\chi_2$ violates CHSH inequality \cite{Cl} maximally and let $\chi_1$ be maximum local. Even in that case Eq.(\ref{x4}) cannot be violated if $\delta_2\,\in [0,1]$ or $\delta_1\,\in [1/2,1]$ or both(see Appendix B). So far we have put forward the cases which cannot assure to show nonbilocality. In this context  we now provide a set of sufficient criteria under which the correlations emerging in a network will definitely exhibit nonbilocality. For this purpose, we consider the subclass  of T states(\ref{p1})
\subsection{Sufficient Criteria for T states}
The inequality (Eq.(\ref{x4})) imposes restrictions on the state parameters of the two copies of T($\tau_i$) state so that nonbilocal correlations are  generated in a network using these restricted copies of T states, if Eq.(\ref{x4}) is violated. Apart from violation of Eq.(\ref{x4}), the variables $\delta_i$, $\epsilon_i$ and $\xi_i$ controlling the state parameters of both local and nonlocal copies should satisfy
\begin{description}
  \item[(i)] $1-\epsilon_i\geq|\xi_i-\delta_i|$
  \item[(ii)] $(\delta_i-\xi_i)^2\geq0$
  \item[(iii)] Depending on the possible values of the state parameters, $\delta_i$, $\epsilon_i$ and $\xi_i$ get further restricted as:
\begin{enumerate}
 \item  $F_i\leq\sqrt{2}+|G_i-H_i|$ when ($c_{1i}\pm c_{2i}$ are of same sign, and $c_{2i}$ is of opposite sign and $c_{3i}>0$) or ($c_{1i}\pm c_{2i}$, and $c_{2i}$ are of same sign and $c_{3i}<0$).
\item  $H_i\leq\sqrt{2}-(F_i+H_i)$  when ($c_{1i}\pm c_{2i}$ are of same sign, and $c_{2i}$ is of opposite sign and $c_{3i}<0$) or ($c_{1i}\pm c_{2i}$, and $c_{2i}$ are of same sign  and $c_{3i}>0$)or ($c_{1i}-c_{2i}$ , $c_{1i}$ are of same sign but $c_{1i}+c_{2i}$ is of opposite sign and $c_{3i}<0$) or ($c_{1i}+c_{2i}$ , $c_{1i}$ are of same sign but $c_{1i}-c_{2i}$ is of opposite sign and $c_{3i}>0$).
\item $G_i\leq\sqrt{2}+|F_i-H_i|$ when ($c_{1i}-c_{2i}$ , $c_{1i}$ are of same sign but $c_{1i}+c_{2i}$ is of opposite sign and $c_{3i}>0$) or ($c_{1i}+c_{2i}$ , $c_{1i}$ are of same sign but $c_{1i}-c_{2i}$ is of opposite sign and $c_{3i}<0$).
\end{enumerate}
where $F_i=\sqrt{1-\epsilon_i+\xi_i-\delta_i}$, $G_i=\sqrt{1-\epsilon_i-\xi_i+\delta_i}$ and $H_i=\sqrt{1+\epsilon_i-\xi_i-\delta_i}$ .
\end{description}
 A particular subclass of T states is the one parameter family of alpha states \cite{ASM}:
$$\rho(\alpha^{'})=\left(
  \begin{array}{cccc}
    \frac{\alpha^{'}}{2} & 0 & 0 &  \frac{\alpha^{'}}{2} \\
    0 &  \frac{1-\alpha^{'}}{2} & 0 & 0 \\
    0 & 0 &  \frac{1-\alpha^{'}}{2} & 0 \\
     \frac{\alpha^{'}}{2} & 0 & 0 &  \frac{\alpha^{'}}{2} \\
  \end{array}
\right)$$
with $\alpha^{'}\in[0,1].$ For this family Eq.(\ref{x4}) takes the form:
\begin{equation}\label{s1}
     \sqrt{\prod_{i=1}^2(2\alpha^{'}_i-1)+\Pi_{i=1}^2\alpha^{'}_i}>1
\end{equation}
$\alpha^{'}_i$ denotes the parameter of the $i$-th copy $(i=1,2)$ of the alpha state. Here $\epsilon_i=1-2(\alpha^{'}_i)^2$, $\delta_i=\xi_i=1-(\alpha^{'}_i)^2-(2\alpha^{'}_i-1)^2$. Clearly, criteria (i) and (ii) are satisfied by $\alpha^{'}_i$. Depending on the signature of $c_{3i}$ some restrictions imposed on $\alpha^{'}_i$ by criteria (iii). If $c_{3i}=2\alpha^{'}_i-1>0,$ i.e., $\alpha^{'}_i>\frac{1}{2}$ then $\alpha^{'}_i$ restricted by criteria (1)of (iii), whereas if $c_{3i}<0$ then criteria (2) restricts the range of $\alpha^{'}_i$. However, one can easily check that no further restriction is imposed on the state parameter other than it should lie in the range $[\frac{1}{2},1]$ in the former case, whereas the range should be $[0,\frac{1}{2}]$ in the latter one. Finally, under the restriction imposed by Eq.(\ref{s1}) the two copies of alpha state can generate nonbilocality only if $\alpha^{'}_i\geq\frac{1}{2}(i=1,2)$ (see FIG.5).
\begin{center}
\begin{figure}[htb]
\centering
\includegraphics[width=2in]{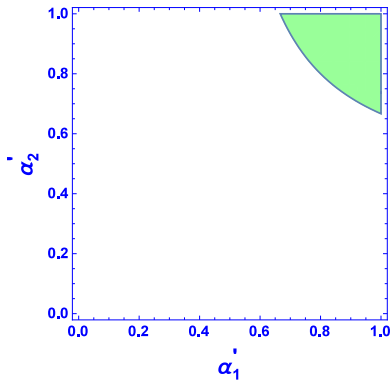}
\caption{\emph{The region of the parameters ($\alpha^{'}_1$, $\alpha^{'}_2$) of the two copies of alpha states which generates nonbilocality is shown here.}}
\end{figure}
\end{center}
\subsection{Criteria for maximal violation of Eq.(\ref{x4})}
 It becomes clear from the symmetry of $\epsilon_i$ and $\xi_i$ (Eq.(\ref{x4})) that T states violate the nonbilocal inequality(Eq.(\ref{x4})) maximally in the $\xi_i=\epsilon_i, ~ (i=1,2)$ plane. The assumption $\epsilon_i$ $=$ $\xi_i, ~ (i=1,2)$ implies $|c_{1i}|=|c_{3i}|$. For $\xi_i$ $=$ $\epsilon_i, ~(i=1,2)$ Eq.(\ref{x4}) becomes:
\begin{equation}\label{x4i}
    \prod_{i=1}^2(1-\delta_i)\leq 1
\end{equation}
\begin{figure}[htb]
\centering
\includegraphics[width=2.2in]{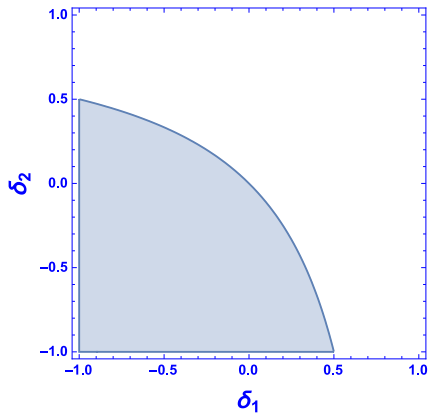}
\caption{\emph{Restriction over $\delta_1$ and $\delta_2$ for generating nonbilocal correlations show that both $\delta_1$  and $\delta_2$ can take max value} $\frac{1}{2}$. Also both of them cannot be positive simultaneously.}
\end{figure}
It becomes clear from Eq.(\ref{x4i}) that under projective measurements, violation of Eq.(\ref{x4i}) in a network is due to $\delta_i$ ($i=1,2$) and hence the locality of the local copy and nonlocal nature of the nonlocal copy are required to ( due to $\delta_1$ and $\delta_2$ respectively) violate this inequality(see FIG.6). We now impose restrictions on the variables $\epsilon_i(=\xi_i),\,\textmd{and}\,\delta_i (i=1,2)$ of local copy and nonlocal copy separately which in turn restrict the parameters of the two copies(local and nonlocal) of the state produced by the two sources $S_i (i=1,2)$ simultaneously. These restricted states suffice to generate nonbilocality in a network.  An obvious restriction is imposed on $\delta_1$ and $\delta_2$(see FIG.6) through the inequality
\begin{equation}\label{x8}
    \prod_{i=1}^2(1-\delta_i)>1.
\end{equation}
For further restrictions, we start with the T states. \textit{When $\tau_1$ local and $\tau_2$ nonlocal}, the restrictions are (for proof, see Appendix. B) :
\begin{description}
\item [C1]: $0\,\leq\,\delta_1\,\leq\,\epsilon_1\,\leq\frac{4\sqrt{2(1-\delta_1)}+5(\delta_1-1)}{2},$ when ($c_{11}\pm c_{21}$ are of same but$c_{21}$ is of opposite sign and $c_{31}>0$) or, ($c_{11}\pm c_{21}$ and $c_{21}$ are of same sign and $c_{31}<0$).
\item[C2]:  $0\,\leq\,\epsilon_1\,\leq\,\delta_1,$ when ($c_{11}- c_{21}$ , $c_{11}$ are of same sign but $c_{11}+ c_{21}$ is of opposite sign and $c_{31}>0$) or, ($c_{11}+c_{21}$, $c_{11}$ are of same sign but $c_{11}- c_{21}$ is of opposite sign and $c_{31}<0$).
\item[C3]: $\delta_2\,\leq\,\epsilon_2\,\leq\frac{4\sqrt{2(1-\delta_2)}+5(\delta_2-1)}{2},$ when ($c_{12}\pm c_{22}$ are of same sign and $c_{22}$ is of opposite sign and $c_{32}>0$) or, ($c_{12}\pm c_{22}$  and$c_{22}$ are of same sign and $c_{32}<0$).
\item[C4]: $\frac{\delta_2-1}{2}\,\leq\,\epsilon_2\,\leq\delta_2,$  when ($c_{12}-c_{22}$ , $c_{12}$ are of same sign but $c_{12}+c_{22}$ is of opposite sign and $c_{32}>0$) or, ($c_{12}+c_{22}$, $c_{12}$ are of same sign but $c_{12}-c_{22}$ is of opposite sign and $c_{32}<0$).
\end{description}
with $\delta_1\in[0,1/2)$ and $\delta_2\in(-1,0)$.\\

\textit{When $\tau_i (i=1,2)$ both nonlocal:}  In this case, sign of $\delta_1$ gives rise to two subcases. If $\delta_1\in[0,1/2)$ then the restrictions on $\tau_1$ gets modified as $$\frac{\delta_1-1}{2}\leq \epsilon_1<0,$$ when ($c_{11},$ $c_{11}-c_{21}$ are of same sign but $c_{11}+c_{21}$ is of opposite sign and $c_{31}>0$) or, ($c_{11}+c_{21}$,$c_{11}$ are of same sign but $c_{11}-c_{21}$ is of opposite sign and $c_{31}<0$), whereas that on $\tau_2$ remain same. If $\delta_1<0$ then the restrictions on $\epsilon_1$ and $\delta_1$ are the same, as those imposed on $\epsilon_2$ and $\delta_2$.

\textit{Conclusion}: In \cite{BRA} the authors showed that there exists quantum advantage in a network under the assumption of source independence. `Quantum advantage' refers to revelation of nonlocal character of correlations (generated in a network) apart from the standard CHSH sense. In this context, we have analyzed the various utilities of the source independence assumption to generate nonlocal correlations in a network. The above discussions clearly justifies that the bilocal assumption is more useful and hence the corresponding Bell-type inequality(Eq.\ref{A3}) is more efficient to demonstrate quantumness of correlations compared to some well-known  preexisting method such as revelation of hidden nonlocality of a state by performing sequential measurements and concepts such as steering nonlocality.  Rest of our paper deals with the restrictions which when imposed on the states generated by the sources in a network (characterized by source independence), suffice to produce quantum advantage. To be specific, we have provided a set of sufficient  criteria for two qubit T states. The related  discussion so far points out the fact that violation of Eq.(\ref{A3}) acts as a sufficient criteria to demonstrate quantum advantage in a network using two qubit states. \\
The bilocal scenarios were introduced in \cite{BRA}. Till then this crucial assumption has been the topic of discussion in few other papers(\cite{BRAN,Fritz,Ross,Ros,Tav,km}). The discussion in this manuscript however reveals the importance of this assumption from the application viewpoint. This in turn points out manifold directions of potential future research. It is clear from the restrictions on the class of Werner states (\ref{x4iii}) that both the copies of the state used in the network (under projective measurements) must be entangled so as to give quantum advantage via the violation of this inequality. But Eq.(\ref{A3}) is only a sufficient condition to get nonbilocal correlations \cite{BRAN}. Intuitively it will be interesting to modify this inequality such that violation of the modified inequality serves not only as a sufficient but also as a necessary condition to give quantum advantage. Also it will be interesting if Alice and Charlie perform more generalized measurements(POVMs).  As generation of nonbilocal correlations is unambiguously more useful in exploiting nonlocal behavior in a quantum system, it may be interesting to find the degrees of relaxation of physical constraints such as measurement independence, determinism required to explain violation of the bilocal inequality Eq.(\ref{A3}) and compare the same with the corresponding degrees of relaxation required for violation of standard Bell-type inequalities\cite{hall2,bp1,bp2}.
\\
\textit{Acknowledgement:}
The authors acknowledge fruitful discussions with SRF A.Sen and S.Karmakar. The author DS acknowledges financial support from DST SERB.

\appendix
\section{}
To derive the bilocal inequality(Eq.\ref{r5}) upto projective measurements  we follow the method used in \cite{bp3}. As already discussed in the main text, in the bilocal network, source $S_1$ sends one copy of X state $\chi_1$(Eq.(\ref{st9})) to Alice and Bob and  source $S_2$ sends a different copy of X state $\chi_2$ to Bob and Charlie. Bob first performs full Bell basis measurement on his two particles. The four outcomes are given by $b^0b^1=00,01,10,11$. He outputs $00,01,10,11$ if he obtains Bell state $|\phi^+\rangle$, $|\phi^-\rangle$, $|\psi^+\rangle$ and $|\psi^-\rangle$ respectively. The final state is shared between Alice and Charlie. Each of these two parties then perform any of two projective measurements. Alice performs projective measurement in any one of two arbitrary directions $\overrightarrow{\alpha_i}$: $A_i=\overrightarrow{\alpha_i}.\overrightarrow{\sigma}(i=1,2)$ where $\overrightarrow{\alpha_i}=(\sin\theta^\alpha_i\cos\phi^\alpha_i,\sin\theta^\alpha_i\sin\phi^\alpha_i,\cos\theta^\alpha_i),\,(i=1,2).$ The set of projective measurements of Charlie are given by $C_i=\overrightarrow{\gamma_i}.\overrightarrow{\sigma}(i=1,2)$ where $\overrightarrow{\gamma_i}=(\sin\theta^\gamma_i\cos\phi^\gamma_i,\sin\theta^\gamma_i\sin\phi^\gamma_i,\cos\theta^\gamma_i),\,(i=1,2)$. Under arbitrary projective measurement settings for Alice and Charlie, the correlation terms $I$ and $J$(Eq.\ref{A3}) are given by:
\begin{widetext}
\begin{equation}\label{f1}
   I= \frac{\Pi_{i=1}^2(\varsigma_i-\kappa_i-\zeta_i+d_i)(\cos\theta_1^\alpha+\cos\theta_2^\alpha)(\cos\theta_1^\gamma+\cos\theta_2^\gamma)}{4}
\end{equation}

\begin{equation}\label{f2}
  J= \frac{4\Pi_{i=1}^2(p_i+q_i)(\sin\theta_1^\alpha\cos\phi_1^\alpha-\sin\theta_2^\alpha\cos\phi_2^\alpha)(\sin\theta_1^\gamma\cos\phi_1^\gamma-\sin\theta_2^\gamma\cos\phi_2^\gamma)}{4}
\end{equation}
\end{widetext}

So the bilocal inequality(Eq.\ref{A3}) takes the form:
 \begin{equation}\label{f3}
  \textbf{B}\leq1.
 \end{equation}
 From Eq.(\ref{f3}) it is clear that $\textbf{B}$ involves eight measurement angles. In order to obtain the bound of the bilocal inequality(Eq.(\ref{r5})) we have to maximize $\textbf{B}$ over these eight measurement angles. Maximizing over $\phi_1^\gamma$, $\phi_2^\gamma$, $\phi_1^\alpha$ and $\phi_2^\alpha$, we get
 \begin{widetext}
 \begin{equation}\label{f4}
  \textbf{B}\leq  \sqrt{\frac{\Pi_{i=1}^2(\varsigma_i-\kappa_i-\zeta_i+d_i)|(\cos\theta_1^\alpha+\cos\theta_2^\alpha)(\cos\theta_1^\gamma+\cos\theta_2^\gamma)|}{4}} +\sqrt{ \frac{4|\Pi_{i=1}^2(p_i+q_i)(\sin\theta_1^\alpha+\sin\theta_2^\alpha)(\sin\theta_1^\gamma+\sin\theta_2^\gamma)|}{4}},
 \end{equation}
 \end{widetext}
 the maximum being obtained for $\phi_1^\gamma$$=$$\phi_1^\alpha$$=$$0$ and $\phi_2^\gamma$$=$$\phi_2^\alpha$$=$$\pi.$ The bound of $\textbf{B}$ given by Eq.(\ref{f4}) involves four measurement angles. Clearly there exists a symmetry in between the measurement angles $\theta_1^\alpha$ and $\theta_2^\alpha$ parameterizing measurement settings of Alice. So maximizing over $\theta_1^\alpha$ and $\theta_2^\alpha$ is equivalent to maximization over any one of them, i.e. we can assume $\theta_1^\alpha$$=$$\theta_2^\alpha$. Similarly a symmetry exists in between the measurement angles $\theta_1^\gamma$ and $\theta_2^\gamma$ parameterizing measurement settings of Charlie and so we can assume $\theta_1^\gamma$$=$$\theta_2^\gamma$. Also there exists symmetry in  between the two pairs $(\theta_1^\alpha,\theta_2^\alpha)$ and $(\theta_1^\gamma,\theta_2^\gamma)$.  So we can replace one pair by the other. Hence in totality, the maximum value can be obtained when all the four measurement angles are equal: $\theta_1^\alpha$$=$$\theta_2^\alpha$$=$$\theta_1^\gamma$$=$$\theta_2^\gamma$$=$$\beta$ (say). Hence
 \begin{widetext}
 $$ \textbf{B}\leq\sqrt{\frac{4\Pi_{i=1}^2(\varsigma_i-\kappa_i-\zeta_i+d_i)\cos^2\beta}{4}}+ \sqrt{16\frac{|\Pi_{i=1}^2(p_i+q_i)|\sin^2\beta}{4}}$$
    $$=\sqrt{\Pi_{i=1}^2(\varsigma_i-\kappa_i-\zeta_i+d_i)}\cos\beta+ \sqrt{4|\Pi_{i=1}^2(p_i+q_i)|}\sin\beta$$
  \begin{equation}\label{f5}
   \leq\sqrt{(\sqrt{\Pi_{i=1}^2(\varsigma_i-\kappa_i-\zeta_i+d_i)})^2+(\sqrt{4|\Pi_{i=1}^2(p_i+q_i)|})^2},
\end{equation}
 \end{widetext}
 where the last inequality is obtained by using the inequality $k_1\cos\theta+k_2\sin\theta\leq\sqrt{k_1^2+k_2^2}$, for any positive value of $k_1$ and $k_2$. This inequality(Eq.(\ref{f5})) ultimately gives the form of the bilocal inequality given by Eq.(\ref{r5}).
\section{}
First we will derive the bilocal inequality Eq.(\ref{x4}). Eqs.(\ref{x3i},\ref{x4i}) give:
\begin{equation}\label{b1}
 (p_i^2+q_i^2)=\frac{1-\epsilon_i}{8}
\end{equation}
\begin{equation}\label{b2}
 (\varsigma_i-\kappa_i-\zeta_i+d_i)^2+4(p_i+q_i)^2=1-\delta_i
\end{equation}
\begin{equation}\label{b3}
  (\varsigma_i-\kappa_i-\zeta_i+d_i)^2+4(p_i-q_i)^2=1-\xi_i
\end{equation}
It is clear from Eqs.(\ref{b1},\ref{b2},\ref{b3}) that
\begin{equation}\label{b3i}
1-\delta_i>|\xi_i-\epsilon_i|.
\end{equation}

Using Eqs.(\ref{b1},\ref{b2},\ref{b3}) we get $\varsigma_i-\kappa_i-\zeta_i+d_i=\pm\sqrt{\frac{1-\delta_i+\epsilon_i-\xi_i}{2}}$ and $2(p_i+q_i)=\pm\sqrt{\frac{1-\delta_i-\epsilon_i+\xi_i}{2}}$. This in turn gives Eq.(\ref{x4}).\\
 Now Eq.(\ref{x4}) can be written as :
\begin{equation}\label{b8}
(\sqrt{\sqrt{\prod_{i=1}^2|P_i-Q_i|}+\sqrt{\prod_{i=1}^2|P_i+Q_i|}}\leq\sqrt{2}
\end{equation}
where $P_i=1-\delta_i$ and $Q_i=\xi_i-\epsilon_i$. Clearly $P_i\geq0$. If $\delta_i\geq0$ then $P_i\in[0,1]$ and if $\delta_i<0$ then $P_i\in[1,2]$. Eq.(\ref{b3i}) implies $P_i>|Q_i|(i=1,2)$. Now we consider $\tau_i(i=1,2)$ to be both local. In that case maximum value of $$(\sqrt{\sqrt{\prod_{i=1}^2|P_i-Q_i|}+\sqrt{\prod_{i=1}^2|P_i+Q_i|}}$$ is $\sqrt{2}$. Hence the correlations produced in a network using two local copies $\tau_i$ cannot violate Eq.(\ref{x4}). Again let $\delta_2<0$ and $\delta_1\geq\frac{1}{2}$. Then the maximum value of $(\sqrt{\sqrt{\prod_{i=1}^2|P_i-Q_i|}+\sqrt{\prod_{i=1}^2|P_i+Q_i|}}$ is $\sqrt{2}$. Hence in this case also Eq.(\ref{x4}) cannot be violated. So if $\delta_i>0(i=1,2)$ or if one of the $\delta_i(i=1,2)$ say $\delta_2<0$ and the other one $\delta_1\geq\frac{1}{2}$ or both then Eq.(\ref{x4}) cannot be violated.

\textit{Proof of the sufficient criterion.}
To prove the sufficient criterion, first we introduce the following variables $A^i$, $B^i$, $C^i$, $D^i$ and $E^i$ such that $A^i = 2\sqrt{2}p_i $, $B^i = 2\sqrt{2}q_i$, $C^i = 2(p_i+q_i)$, $D^i = 2(p_i - q_i)$, $E^i = (\varsigma_i - \kappa_i - \zeta_i + d_i)$ $(i = 1, 2)$. Then Eqs.(20,21,22) can be written in the form
\begin{equation}\label{A1}
    (A^i)^2 + (B^i)^2 = 1 -\epsilon_i
\end{equation}
\begin{equation}\label{A2}
   (E^i)^2 + (C^i)^2 = 1 - \delta_i
\end{equation}
\begin{equation}\label{p3}
    (E^i)^2 + (D^i)^2 = 1 - \xi_i
\end{equation}
Hence $(C^i)^2 = \frac{1 + \xi_i - \epsilon_i - \delta_i}{2}$, $ (D^i)^2 = \frac{1 - \xi_i - \epsilon_i + \delta_i}{2}$ and $(E^i)^2 = \frac{1 - \xi_i + \epsilon_i - \delta_i}{2}$. State conditions impose the constraints $p_i^2 \leq {\varsigma_i}{d_i}$ and $q_i^2 \leq {\kappa_i}{\zeta_i}$. Since $\varsigma_i = d_i$, $\kappa_i = \zeta_i$ (for T states), $2(\varsigma_i + \zeta_i) = 1$ and $E^i = 2(\varsigma_i - \zeta_i )$, the above constraints reduces to
\begin{equation}\label{A4}
  \frac{ -(1 + E^i)}{\sqrt{2}} \leq A^i \leq   \frac{ (1 + E^i)}{\sqrt{2}}
\end{equation}
and
\begin{equation}\label{A5}
   \frac{ -(1 - E^i)}{\sqrt{2}} \leq B^i \leq   \frac{ (1 - E^i)}{\sqrt{2}}.
\end{equation}
Again $A^i = \frac{C^i + D^i}{\sqrt{2}}$ and $B^i = \frac{C^i - D^i}{\sqrt{2}}$, so the above equations get modified as
\begin{equation}\label{A6}
  \frac{ -(1 + E^i)}{\sqrt{2}} \leq \frac{C^i + D^i}{\sqrt{2}} \leq   \frac{ (1 + E^i)}{\sqrt{2}}
\end{equation}
and
\begin{equation}\label{A7}
   \frac{ -(1 - E^i)}{\sqrt{2}} \leq \frac{C^i - D^i}{\sqrt{2}} \leq   \frac{ (1 - E^i)}{\sqrt{2}}.
\end{equation}
For T states $\varsigma_i-\kappa_i-\zeta_i+d_i=2c_{3i}$, $p_i-q_i=-2c_{2i}$ and $p_i+q_i=-2c_{1i}$
There are $ 16 $ possible different cases corresponding to the different sign of $A^i$, $B^i$, $C^i$, $D^i$ and $E^i$ which ultimately give rise to the following four pairs of inequalities:

 \begin{itemize}
\item  1a.) $0\leq\frac{\sqrt{1 + \xi_i - \epsilon_i - \delta_i} + \sqrt{1 - \xi_i - \epsilon_i + \delta_i}}{2}\leq\frac{\sqrt{2} + \sqrt{1 - \xi_i + \epsilon_i - \delta_i}}{2}$ \\
1b.) $0\leq\frac{\sqrt{1 + \xi_i - \epsilon_i - \delta_i} - \sqrt{1 - \xi_i - \epsilon_i + \delta_i}}{2}\leq\frac{\sqrt{2} - \sqrt{1 - \xi_i + \epsilon_i - \delta_i}}{2}$ \\
when($c_{1i}\pm c_{2i}$ are of same sign and $c_{2i}$ is of opposite sign and $c_{3i}>0$) or ($c_{1i}\pm c_{2i}$  and $c_{2i}$ are of same sign and $c_{3i}<0$).
 \item 2a.) $0 \leq \frac{\sqrt{1 + \xi_i - \epsilon_i - \delta_i} - \sqrt{1 - \xi_i - \epsilon_i + \delta_i}}{2} \leq \frac{\sqrt{2} + \sqrt{1 - \xi_i + \epsilon_i - \delta_i}}{2}$\\
         2b.) $0\leq\frac{\sqrt{1 + \xi_i - \epsilon_i - \delta_i} + \sqrt{1 - \xi_i - \epsilon_i + \delta_i}}{2}\leq\frac{\sqrt{2} - \sqrt{1 - \xi_i + \epsilon_i - \delta_i}}{2}$\\
       when($c_{1i}\pm c_{2i}$ are of same sign and $c_{2i}$ is of opposite sign and $c_{3i}<0$) or ($c_{1i}\pm c_{2i}$  and $c_{2i}$ are of same sign and $c_{3i}>0$).
   \item 3a.) $0\leq\frac{\sqrt{1 - \xi_i - \epsilon_i + \delta_i} - \sqrt{1 + \xi_i - \epsilon_i - \delta_i}}{2}\leq\frac{\sqrt{2} + \sqrt{1 - \xi_i + \epsilon_i - \delta_i}}{2}$\\
         3b.) $0\leq\frac{\sqrt{1 + \xi_i - \epsilon_i - \delta_i} + \sqrt{1 - \xi_i - \epsilon_i + \delta_i}}{2}\leq\frac{\sqrt{2} - \sqrt{1 - \xi_i + \epsilon_i - \delta_i}}{2}$\\
         when ($c_{1i}-c_{2i}$ , $c_{1i}$ are of same sign but $c_{1i}+c_{2i}$ is of opposite sign and $c_{3i}<0$) or ($c_{1i}+c_{2i}$,$c_{1i}$ are of same sign but $c_{1i}-c_{2i}$ is of opposite sign and $c_{3i}>0$).
   \item 4a.) $0\leq\frac{\sqrt{1 + \xi_i - \epsilon_i - \delta_i} + \sqrt{1 - \xi_i - \epsilon_i + \delta_i}}{2}\leq\frac{\sqrt{2} + \sqrt{1 - \xi_i + \epsilon_i - \delta_i}}{2}$\\
         4b.) $0\leq\frac{\sqrt{1 - \xi_i - \epsilon_i + \delta_i} - \sqrt{1 + \xi_i - \epsilon_i - \delta_i}}{2}\leq\frac{\sqrt{2} - \sqrt{1 - \xi_i + \epsilon_i - \delta_i}}{2}$\\
        when ($c_{1i}-c_{2i}$ , $c_{1i}$ are of same sign but $c_{1i}+c_{2i}$ is of opposite sign and $c_{3i}>0$) or ($c_{1i}+c_{2i}$,$c_{1i}$ are of same sign but $c_{1i}-c_{2i}$ is of opposite sign and $c_{3i}<0$).
 \end{itemize}
Throughout the discussion below we consider $\tau_2$ as the nonlocal copy. Now for the other copy $\tau_1$, first we consider it to be local.

\textit{Different cases of the local copy($\tau_1$):} under the assumption of $\xi_1 = \epsilon_1$. First we consider the pair of inequalities 1a and 1b. Positivity of $(1 + \delta_1 - 2\epsilon_1 )$ implies
  \begin{equation}\label{A7'}
    \epsilon_1 \leq \frac{1 + \delta_1}{2}.
\end{equation}
inequalities of 1a are obvious for a local copy(i.e., for $\delta_1$ and $\xi_1 \in [0 , 1]$). Under the assumption $\xi_1 = \epsilon_1$, first and second inequalities of (1b) becomes
 \begin{equation}\label{A8}
 \sqrt{1 - \delta_1} - \sqrt{1 + \delta_1 - 2\epsilon_1} \geq 0
 \end{equation}
 and
 \begin{equation}\label{A9}
    \sqrt{1 - \delta_1} - \sqrt{1 + \delta_1 - 2\epsilon_1} \leq \sqrt{2} - \sqrt{1 - \delta_1}
 \end{equation}
 respectively. Clearly, Eq.(\ref{A8}) and \ Eq.(\ref{A9}) implies $\delta_1 \leq \epsilon_1$ and
 \begin{equation}\label{A10}
   \sqrt{\frac{1 + \delta_1 - 2\epsilon_1}{2}} \geq \sqrt{2(1 - \delta_1)} - 1
 \end{equation}
 respectively.
 The above relation is obvious if $\delta_1 \geq \frac{1}{2}$. By taking $\delta_1 \leq \frac{1}{2}$, the above relation(Eq.(\ref{A10})) gets modified as
 \begin{equation}\label{A11}
    \epsilon_1 \leq \frac{5(\delta_1 - 1) + 4\sqrt{2(1 - \delta_1)}}{2}
 \end{equation}
 Hence, for $\delta_1 \le \frac{1}{2},$ Eq.(\ref{A7'}) and Eq.(\ref{A11}) implies
 \begin{equation}\label{A12}
   \epsilon_1 \leq \min\{\frac{5(\delta_1 - 1) + 4\sqrt{2(1 - \delta_1)}}{2} , \frac{1 + \delta_1}{2} \}.
 \end{equation}
But $ \frac{5(\delta_1 - 1) + 4\sqrt{2(1 - \delta_1)}}{2} \leq  \frac{1 + \delta_1}{2}. $ Hence for a local copy  $0\leq\delta_1 \leq \epsilon_1 \leq \frac{5(\delta_1 - 1) + 4\sqrt{2(1 - \delta_1)}}{2}$ when $\delta_1 \le \frac{1}{2}$. Following the same procedure as for 1a and 1b, for the last pair (i.e 4a and 4b) we have $0 \leq \epsilon_1 \leq \delta_1$. Now we consider the pair (2a, 2b); similarly as in the previous cases, from the second inequality of (2b) we get

\begin{equation}\label{A13}
\sqrt{\frac{1 + \delta_1 - 2\epsilon_1}{2}} \leq  1 - \sqrt{2(1 - \delta_1)}.
\end{equation}
The above inequality is satisfied only when $\delta_1 \geq \frac{1}{2}.$ Again from Eq.(\ref{x4i})it is clear that the inequality holds only when $\delta_1 < \frac{1}{2}.$ Hence nonbilocal correlation cannot be obtained if one uses a local state whose parameters are restricted by this pair of inequalities(2a and 2b) as in a network under projective measurements and (non)bilocal inequality considered here. Same analysis holds good for the pair (3a) and (3b). \\

\textit{Different cases of the nonlocal copy($\tau_2$):} Here, $\delta_2$ and/or $\epsilon$ $\in [-1 , 0)$. But as already discussed before, when $\tau_1$ is local($\delta_1>0$) then if $\delta_2>0$, Eq.(\ref{x4}) cannot be violated.  Hence $\delta_2<0$ is a sufficient condition for violation of Eq.(\ref{x4}). For (1a,1b) case, first inequality of 1a imposes obvious restriction. Again positivity of $(1 + \delta_2 - 2\epsilon_2)$ and second inequality of (1a) i.e., $\frac{\sqrt{1 + \xi_2 - \epsilon_2 - \delta_2} - \sqrt{1 - \xi_2 - \epsilon_2 + \delta_2}}{2}\leq\frac{\sqrt{2} - \sqrt{1 - \xi_2 + \epsilon_2 - \delta_2}}{2}$ together give:
\begin{equation}\label{A14}
   \frac{\delta_2 - 1}{2} \leq \epsilon_2 \leq \frac{1 + \delta_2}{2}.
\end{equation}
Also from first and second inequality of (2b) implies
\begin{equation}\label{A15}
    \delta_2 \leq \epsilon_2 \leq \frac{5(\delta_2 - 1) + 4\sqrt{2(1 - \delta_2)}}{2}\, \textmd{for} \delta_2 < 0.
\end{equation}
It is easy to check that the restriction imposed on $\epsilon_2$ in Eq.(\ref{A15}) is more stronger than in Eq.(\ref{A14}). Hence, Eq.(\ref{A14}) and Eq.(\ref{A15}) together imply $\delta_2 \leq \epsilon_2 \leq \frac{5(\delta_2 - 1) + 4\sqrt{2(1 - \delta_2)}}{2}$ for $\delta_2 < \frac{1}{2}$. Following the same procedure as in the previous case, for the last case (4a, 4b) we have $ \frac{\delta_2 - 1}{2} \leq \epsilon_2 \leq \delta_2 $. Now we consider (2a,2b) as a nonlocal copy, then from the first inequality of (2a) and second inequality of (2b) we get
\begin{equation}\label{A16}
    \delta_2 \leq \epsilon_2
\end{equation}
and
\begin{equation}\label{A17}
   \sqrt{\frac{1 + \delta_2 - 2\epsilon_2}{2}} \leq  1 - \sqrt{2(1 - \delta_2)}.
\end{equation}
respectively.
But the last Eq.(\ref{A17}) holds only when $\delta_2 > \frac{1}{2}$ and since  $\delta_2 \leq \epsilon_2$, So Eq.(\ref{A16}) and Eq.(\ref{A17}) simultaneously hold only when $\frac{1}{2}<\delta_2 <\epsilon_2$. But this is not possible for a nonlocal copy.

Similarly as in local copy nonbilocal correlation cannot be obtained when one use (2a,2b) and (3a,3b) as a nonlocal copy. Hence in order to get nonbilocal correlation (1a,1b) and(4a,4b) are the only valid pairs of inequalities to impose restrictions on both local and nonlocal copies($\tau_1$ and $\tau_2$ respectively) which along with restriction given by Eq.(\ref{x4i}), suffice to violate Eq.(\ref{x4}).
Next we consider both $\tau_i(i=1,2)$ to be nonlocal. Firstly we put another restriction on $\tau_1$ by considering $\delta_1>0$. Under these restrictions (4a,4b) is the only valid pair of inequality which imposes the restriction: $\frac{\delta_1-1}{2}\leq\epsilon_1<0$. But if $\delta_1<0$, then we get the same restrictions as that for $\delta_2$ and $\epsilon_2$.
\section{}
Maximal bound for violation of CHSH inequality by the filtered state X state(Eq.(\ref{r4})) is enlisted in the following table.
\begin{widetext}
\begin{table}[htp]
\begin{center}
\begin{tabular}{|c|c|c|}
\hline
  State Parameter & Bound For Violation of CHSH\\
  \hline
   $pq>0$ & $\textmd{Max}[8\frac{\sqrt{2(p^2+q^2)(\lambda_1\lambda_2)^2}}{N_1},\frac{2\sqrt{4(p+q)^2(\lambda_1\lambda_2)^2+(d-c\lambda_2^2-\kappa\lambda_1^2+\varsigma(\lambda_1\lambda_2)^2)^2}}{N_1}]$\\

 \hline
    $pq<0$ & $\textmd{Max}[8\frac{\sqrt{2(p^2+q^2)(\lambda_1\lambda_2)^2}}{N_1},\frac{2\sqrt{4(p-q)^2(\lambda_1\lambda_2)^2+(d-c\lambda_2^2-\kappa\lambda_1^2+\varsigma(\lambda_1\lambda_2)^2)^2}}{N_1}]$\\
\hline
\end{tabular}\\
\end{center}
\caption{The bounds for violation of CHSH inequality by a filtered X state(Eq.(\ref{r4}))  are listed here. The bound differs depending on the signature of $pq$. Here $N_1=\varsigma\lambda_1^2\lambda_2^2+\kappa\lambda_1^2+\zeta\lambda_2^2+d$.}
\label{table1}
\end{table}
\end{widetext}

\begin{thebibliography}{1}
\bibitem{BRA} C. Branciard, N. Gisin, and S. Pironio, Phys. Rev. Lett. \textbf{104},170401 (2010).
\bibitem{BRAN} C. Branciard, D. Rosset,N. Gisin and S. Pironio, Phys. Rev. A \textbf{85},032119 (2012).
\bibitem{Ross} D. Rosset, C. Branciard, N. Gisin, and Y. C. Liang, New J. Phys. \textbf{15} 053025 (2013).
\bibitem{Fritz} T. Fritz New J. Phys. \textbf{14} 103001 (2012)
\bibitem{Ros} Cyril Branciard et al., Phys. Rev. Lett. \textbf{109}, 100401 (2012).
\bibitem{Tav} A. Tavakoli, P. Skrzypczyk, D. Cavalcanti and A. Acin, Phys. Rev. A \textbf{90}, 062109 (2014) .
\bibitem{km} K. Mukherjee, B. Paul and D. Sarkar, Quantum Inf Process. \textbf{14}, 2025 (2015).
\bibitem{Bell} J. S. Bell, Physics \textbf{1}, 195 (1964).
\bibitem{Pironio} S. Pironio, A. Acin, S. Massar, A. B. de la Giroday, D. N. Matsukevich,P. Maunz, S. Olmschenk, D. Hayes, L. Luo, T. A.Manning, and C. Monroe, Nature \textbf{464}, 1021 (2010).
\bibitem{Colbeck} R. Colbeck and A. Kent, Journal of Physics A: Mathematical and Theoretical \textbf{44}, 095305 (2011).
\bibitem{Acin}A. Acin, N. Brunner, N. Gisin, S. Massar, S. Pironio, and V. Scarani, Phys. Rev. Lett. \textbf{98}, 230501 (2007).
\bibitem{Mayer}D. Mayers and A. Yao,\,in Proceedings of the 39th IEEE Symposiumon Foundations of Computer Science (IEEE Computer Society, Los Alamitos CA,USA,1998)p.503.
\bibitem{Cleve} R. Cleve and H. Buhrman, Phys. Rev. A \textbf{56}, 1201 (1997).
\bibitem{Bancal} J.D. Bancal, N. Gisin, Y.-C. Liang, and S. Pironio,Phys. Rev. Lett. \textbf{106}, 250404 (2011).
\bibitem{Bar} C.-E. Bardyn, T. C. H. Liew, S. Massar, M. McKague, and  V. Scarani, Phys. Rev. A \textbf{80}, 062327 (2009).
\bibitem{Mck} M. McKague, M. Mosca, arXiv:1006.0150v1 [quant-ph] (2010).
\bibitem{San} N. Sangouard, C. Simon, H. de Riedmatten, and N. Gisin, Rev. Mod. Phys. \textbf{83}, 33 (2011).
\bibitem{HAM} K. Hammerer, A. S. Sørensen, and E. S. Polzik, Rev. Mod. Phys. \textbf{82}, 1041 (2010).
\bibitem{Zuk} M. Zukowski et al., Phys. Rev. Lett. \textbf{71}, 4287 (1993).
\bibitem{Cl} J.F. Clauser, M.A. Horne, A. Shimony, and R.A. Holt. Phys. Rev.Lett. \textbf{23}, 880 (1969).
\bibitem{Wer} R.F. Werner, Phys. Rev. A \textbf{40}, 4277 (1989).
\bibitem{Barr} A. Ac´in, N. Gisin, and B. Toner, Phys. Rev. A \textbf{73}, 062105 (2006).
\bibitem{pop} S. Popescu, Phys. Rev. Lett. \textbf{74}, 2619 (1995).
\bibitem{mas1} L. Masanes, Phys. Rev. Lett. \textbf{97}, 050503 (2006).
\bibitem{mas2} Y.C. Liang, L. Masanes and D. Rosset, Phys. Rev. A \textbf{86}, 052115 (2012).
\bibitem{Hir} F. Hirsch, M. T. Quintino, J. Bowles and  N. Brunner, Phys. Rev. Lett. \textbf{111}, 160402 (2013).
\bibitem{Wise} H. M. Wiseman, S. J. Jones and A. C. Doherty, Phys.Rev. Lett. \textbf{98}, 140402 (2007).
\bibitem{nat} D. J. Saunders, S. J. Jones, H. M. Wiseman and G. J.Pryde, Nature. Phys. \textbf{6}, 845 (2010).
\bibitem{tyu} T. Yu and J. H. Eberly, Quantum Inf. Comput. \textbf{7}, 459 (2007).
\bibitem{ali} Mazhar Ali, A.R.P. Rau and G. Alber, Phys. Rev. A \textbf{81}, 042105 (2010).
\bibitem{rho} R. Horodecki and M. Horodecki, Phys. Rev. A \textbf{54}, 1838 (1996).
\bibitem{BEO} B. Bellomo, R. Lo Franco and G Compagno, Adv. Sci. Lett. \textbf{2}, 459 (2009).
\bibitem{deo} J. Maziero,et al., Phys. Rev. A \textbf{81},022116(2010).
\bibitem{CHI} A. Chiuri, G. Vallone, M. Paternostro, P. Mataloni, Phys. Rev. A \textbf{84}, 020304(R) (2011).
\bibitem{LDI} L. Di Carlo et al., Nature \textbf{460}, 240 (2009)
\bibitem{PET} N. A. Peters, J. B. Altepeter, D. Branning, E. R. Jeffrey, T.C. Wei and P. G. Kwiat, Phys. Rev. Lett. \textbf{92}, 133601 (2004).
\bibitem{PRA} J. S. Pratt, Phys. Rev. Lett. \textbf{93}, 237205 (2004).
\bibitem{BOS} S. Bose, I. Fuentes-Guridi, P. L. Knight, V. Vedral, Phys. Rev. Lett. \textbf{87}, 050401 (2001).
\bibitem{WAN} J. Wang, H. Batelaan, J. Podany, A. F. Starace, J. Phys. B, At. Mol. Opt. Phys. \textbf{39}, 4343 (2006).
\bibitem{HAG} E. Hagley, et al., Phys. Rev. Lett. \textbf{79}, 1 (1997).
\bibitem{Ceo} S. Bose, I. Fuentes-Guridi, P. L. Knight, and V. Vedral,Phys. Rev. Lett. \textbf{87}, 050401 (2001).
\bibitem{OSB} T. J. Osborne and M. A. Nielsen, Phys. Rev. A \textbf{66}, 032110 (2002).
\bibitem{OST} A. Osterloh, Luigi Amico, G. Falci and R. Fazio, Nature \textbf{416}, 608 (2002).
\bibitem{FAN} F. F. Fanchini, T.Werlang, C. A. Brasil, L. G. E. Arruda, and A. O. Caldeira, Phys. Rev. A \textbf{81}, 052107 (2010).
\bibitem{DIL} R. Dillenschneider, Phys. Rev. B \textbf{78}, 224413 (2008).
\bibitem{SAR} M. S. Sarandy, Phys. Rev. A \textbf{80}, 022108 (2009).
\bibitem{WER} T. Werlang, C. Trippe, G. A. P. Ribeiro, and G. Rigolin, Phys. Rev. Lett. \textbf{105}, 095702 (2010).
\bibitem{CIL} L. Ciliberti, R. Rossignoli, and N. Canosa,Phys. Rev. A \textbf{82}, 042316 (2010).
\bibitem{TYU} T. Yu and J. H. Eberly, Phys. Rev. Lett. \textbf{93}, 140404 (2004); Science \textbf{323}, 598 (2009).
\bibitem{VIN} S. Vinjanampathy and A. R. P. Rau, Phys. Rev. A \textbf{82}, 032336 (2010).
\bibitem{zuk1} Marek Zukowski, Arijit Dutta, Zhi Yin, Phys. Rev. A \textbf{91}, 032107 (2015).
\bibitem{HOR} R. Horodecki, P. Horodecki, and M. Horodecki, Phys. Lett. A \textbf{200}, 340 (1995).
\bibitem{Pir} S. Pironio, J. Math. Phys. \textbf{46}, 062112 (2005).
\bibitem{bnn} C. H. Bennett, D. P. DiVicenzo, J. A. Smolin, and W.K. Wootters, Phys. Rev. A \textbf{54}, 3824 (1996).
\bibitem{woo} W. K. Wootters, Phys. Rev. Lett. \textbf{80}, 2245 (1998).
\bibitem{vid} G. Vidal and R. F. Werner, Phys. Rev. A \textbf{65}, 032314 (2002).
\bibitem{ASM} A. Ai-Qasimi, D. F. V. James, arXiv:1007.1814v1 [quant-ph] (2010).
\bibitem{hall2} M. J. W. Hall, Phys. Rev. A \textbf{84}, 022102 (2011).
\bibitem{bp1}B. Paul, K. Mukherjee and D. Sarkar, Phys. Rev. A \textbf{88}, 014104 (2013).
\bibitem{bp2}B. Paul, K. Mukherjee and D. Sarkar, Quantum Inf Process \textbf{13}, 1687–1699 (2014).
\bibitem{bp3} K. Mukherjee, B. Paul, D. Sarkar, J. Phys. A: Math. Theor. \textbf{48}, 465302 (2015).



\end{thebibliography}
\end{document}